\documentclass[fdp,fleqn]{w-art}
\usepackage{times}
\usepackage{w-thm}
\usepackage[]{graphicx}
\usepackage{epstopdf}
\usepackage{amssymb}

\def\ket{\rangle}
\def\bra{\langle}
\def\pard{\partial}

\def\epth{\epsilon_\textrm{th}}

\def\epA{\epsilon_\textrm{A}}
\def\zB{\bar{z}_\textrm{B}}
\def\zL{\bar{z}_\textrm{L}}

\begin{document}
%%    The information for the title page will be placed between
%%    \begin{document} and \maketitle. The order of most entries
%%    is determined by the class file and can not be changed by
%%    rearranging them. The maketitle command follows after the
%%    abstract.
%%
%%    Most of the following commands will be completed by the publisher.
%%
%%    The copyrightyear is defined in the .clo file as the first argument
%%    of the copyrightinfo command. If the copyrightyear differs from that
%%    value it might be adjusted by the following definition:
%%
%% \renewcommand{\copyrightyear}{2003}% uncomment to change the copyrightyear.
%%
\DOIsuffix{theDOIsuffix}
%%
%% issueinfo for header and copyright line
\Volume{55}
\Issue{1}
\Month{01}
\Year{2007}
%%
%%    First and last pagenumber of the article. If the option
%%    'autolastpage' is set (default) the second argument may be left empty.
\pagespan{3}{}
%%
%%    Dates will be filled in by the publisher. The 'reviseddate' and
%%    'dateposted' (Published online) entry may be left empty.
%\Receiveddate{15 November 2007}
%\Reviseddate{30 November 2007}
%\Accepteddate{2 December 2007}
%\Dateposted{3 December 2007}
%%

%% \pretitle{Editor's Choice}

%% We have a short and a long form for the title. The short form
%% (optional argument) goes into the running head.

\title[Bound state de-stabilization and non-Markovian decay]{Amplification of non-Markovian decay due to bound state absorption into continuum}

%% Please do not enter footnotes or \inst{}-notes into the optional
%% argument of the author command. The optional argument will go into
%% the header.  If there is only one address the marker \inst{x} may be
%% omitted.

%% Information for the first author.
\author[S. Garmon]{Savannah Garmon\inst{1,}%
  \footnote{Corresponding author\quad E-mail:~\textsf{sgarmon@chem.utoronto.ca}, 
            Phone: +1\,416\,946\,7466, 
            Fax: +1\,416\,978\,8775}}
\address[\inst{1}]{Chemical Physics Theory Group, Department of Chemistry, University of Toronto,
			80 St. George Street, Toronto, Ontario M5S 3H6, Canada}
%%
%%    Information for the second author
\author[T. Petrosky]{Tomio Petrosky\inst{2,}%\footnote{Second author footnote.} 
										}
\address[\inst{2}]{Center for Complex Quantum Systems, University of Texas at Austin,
			1 University Station, C1609, Austin, Texas 78712}
%%
%%    Information for the third author
\author[Y. Nikulina]{Lena Simine\inst{1,}
										}
\author[D. Segal]{Dvira Segal\inst{1,}%\footnote{Third author footnote.}
										}
%%
%%    \dedicatory{This is a dedicatory.}
\begin{abstract}
It is known that quantum systems yield non-exponential (power law) decay on long time scales, associated with continuum threshold effects contributing to the survival probability for a prepared initial state.
For an open quantum system consisting of a discrete state coupled to continuum, we study the case in which a discrete bound state of the full Hamiltonian approaches the energy continuum as the system parameters are varied.  We find in this case that at least two regions exist yielding qualitatively different power law decay behaviors; we term these the long time `near zone' and long time `far zone.'  In the near zone the survival probability falls off according to a $t^{-1}$ power law, and in the far zone it falls off as 
$t^{-3}$.  We show that the timescale $T_Q$ separating these two regions is inversely related to the gap between the discrete bound state energy and the continuum threshold.  In the case that the bound state is absorbed into the continuum and vanishes, then the time scale $T_Q$ diverges and the survival probability follows the $t^{-1}$ power law even on asymptotic scales.  Conversely, one could study the case of an anti-bound state approaching the threshold before being ejected from the continuum to form a bound state.  Again the $t^{-1}$ power law dominates precisely at the point of ejection.
\end{abstract}
%% maketitle must follow the abstract.
\maketitle                   % Produces the title.

%% If there is not enough space inside the running head
%% for all authors including the title you may provide
%% the leftmark in one of the following three forms:

%% \renewcommand{\leftmark}
%% {F. Author: A short title}

%% \renewcommand{\leftmark}
%% {F. Author and S. Author: A short title}

%% \renewcommand{\leftmark}
%% {F. Author et al.: A short title}

%% \tableofcontents  % Produces the table of contents.
\section{Introduction}\label{sec:intro}

For decades it has been known that deviations from exponential decay in quantum systems exist under quite general conditions, both on extremely short and very long time scales \cite{RGWinter,Sudarshan,LFondaEtAl,Ch9_review}.
The short time scale deviations typically give rise to parabolic decay; rapidly repeated measurements of the system on this timescale may give rise to either the quantum Zeno or quantum anti-Zeno effects, resulting in decelerated or accelerated decay, respectively \cite{Sudarshan,Sudarshan_QZ,Kof&Kur00}.  
However, these effects typically occur on quite short timescales, before giving way to the more familiar exponential decay, and hence experimental study was elusive for several decades; it is only relatively recently that these effects were experimentally verified \cite{short_time_expt,zeno_expt}.

Meanwhile, it was Khalfin \cite{Khalfin}, relying on the Paley-Wiener theorem \cite{PW-thm}, who showed that a quantum system that is bounded below necessarily gives rise to non-exponential decay on asymptotic timescales; an alternative proof that does not rely on the Paley-Wiener theorem was provided by Hack \cite{Hack}.  However, the timescales on which the non-exponential effects manifest typically exceed several lifetimes, after which the survival probability is usually quite depleted, again posing an experimental challenge.  It is therefore quite recently this effect has been verified experimentally in the luminescence decay properties of dissolved organic materials following laser-excitation \cite{long_time_expt}.  The recent advances in the fabrication of nano-scale devices and cold atom trapping provide an optimistic outlook for deeper studies in this direction, particularly given the previous success in detecting non-exponential decay at short times for sodium atoms initially trapped in an accelerating optical potential \cite{short_time_expt}.

The long-time deviation, which typically takes the form of an inverse power law, is intimately connected with the properties of the continuum and particularly the threshold.  
As an example, for a particle initially trapped in a shallow inner potential well it has been shown that as one varies the strength of a repulsive outer inverse square potential, the exponent of the power law varies in a predictable manner \cite{MMS08}.  In another study, it has been shown for a radial potential that for the $s$-wave component of the wave function as one brings the energy of an initially prepared unstable state (a resonant state initially localized in the inner well) close to the threshold that the exponential decay can be completely suppressed, resulting in power law decay on all time scales \cite{JMSST05}.
A similar effect is demonstrated in Ref. \cite{GV06} as a resonant state approaches the threshold in a semiconductor double barrier quantum structure.

None of the three above studies, however, consider the effect of bound states on the power law decay.  A semi-infinite tight-binding chain with an endpoint impurity is considered in Ref. \cite{LonghiPRL06}, in which case exponential decay is again suppressed, however, in this case the effect is associated with the appearance of a bound state.
%%%ADDED IN V2:
The same effect is observed in an equivalent model in Ref. \cite{thesis}, as well as in a semi-infinite chain with a two-level endpoint impurity in Ref. \cite{DBP08}.  In the latter two studies, an intensification of the power law decay is associated with the appearance or dissolution of a bound state (i.e. a transition from a localized state to a virtual state in the language of Ref. \cite{DBP08}).
We also note that the effect of a bound state near threshold on the non-exponential decay is considered from a more mathematical perspective in Ref. \cite{DJN09}.  However, the analysis presented in that work is not rooted in any specific physical Hamiltonian but rather the authors place conditions on the spectrum itself.  For example, they assume that an unperturbed bound state appearing directly at threshold cannot result in a bound state shifted below the threshold under the influence of the interaction with the continuum.  However, this assumption is contradicted by both of the physical models considered in this paper, as well those studied in Refs. \cite{LonghiPRL06,thesis,DBP08}.

In the following study, we consider more closely the role that the bound state plays in relation to the non-exponential decay in open quantum systems.  In particular, we show that in the case that a bound state approaches the energy continuum, there are two different temporal regions in which the leading term in the power law expansion takes on different exponents; we term these the long time ``near zone'' and the long time ``far zone.''   
%%%%
%%%%
\footnote{This naming convention is chosen in analogy with the spatial dependence in the description of the photon dressing for a two-level atom; in that case there is the near zone or Van der Waals zone in which the photon cloud falls off as $x^{-6}$ and the far zone or Casimir-Polder zone in which it falls off as $x^{-7}$.  Finally there is a characteristic length separating these two zones, fixed by the energy gap between the two levels \cite{Passante_book}.
}  %%%
%%%%
The power law decay in the near zone is enhanced relative to that in the far zone and the time scale $T_Q$ that separates the two zones is inversely related to the gap between the bound state energy and the continuum threshold.  Hence as the bound state approaches the threshold the time scale $T_Q$ grows larger, resulting in the amplified near zone decay extending over longer and longer time scales.  Finally, in the case that bound state is absorbed into the continuum at the threshold (meaning that the bound state is destabilized as it vanishes from the diagonalized Hamiltonian spectrum) the time scale $T_Q$ diverges and the near zone description becomes the fully asymptotic large time behavior.

On the other side of this transition, an anti-bound state forms in the second Riemann sheet, meanwhile the long time far zone behavior is restored in the asymptotic limit.  Hence, one could just as well view this transition in reverse: an anti-bound state approaches the energy continuum, and the power law decay is amplified at precisely the point that the anti-bound state touches the continuum edge before being ejected from the continuum as a bound state.
%%%ADDED IN V2:
Indeed, the power law decay is easier to study in the latter case as neither a bound nor a resonance appear in the spectrum (at least, for the model that we consider below), leaving the power law as the most pronounced effect in the time evolution  \cite{LonghiPRL06,thesis,DBP08}.

In Sec. \ref{sec:gen} below we present our argument that the power law decay is amplified as the bound state approaches the continuum threshold (or the edge of the conduction band in solid state terms) and in Sec. \ref{sec:gen.bound.states} we discuss two cases for open quantum systems: in one case the bound state is prevented from reaching the continuum threshold due to a divergent van Hove singularity in the density of states while in the other the form of the interaction potential modifies the system properties and allows the bound state to reach the continuum edge, resulting in a divergence for the timescale $T_Q$.  In Sec. \ref{sec:mod} we look at two model realizations of these effects: in Sec. \ref{sec:modI} we study an infinitely long quantum dot superlattice with a side-coupled impurity (Model I); in this case the van Hove singularity prevents the bound state from destabilizing at the continuum edge.  Meanwhile in Sec. \ref{sec:modII} we present the case of a similar model consisting of a semi-infinite chain with an endpoint impurity (Model II); here the bound state may reach the continuum edge, resulting in the divergence of the timescale $T_Q$.  We make our concluding remarks in Sec. \ref{sec:conc}.

%%%%%%%%%%%%%%%%%
%%%%%%%SECTION BREAK
%%%%%%%%%%%%%%%%%

\section{Bound state de-stabilization and non-Markovian decay}\label{sec:gen}

Here we consider an open quantum system with a generic Hamiltonian of the form 
\begin{equation}
H =
\epsilon_d d^\dag d
% + \int_{\epth^-}^{\epth^+} dk \, \epsilon_k \, c_k^\dag c_{k}
 + \int dk \, \epsilon_k \, c_k^\dag c_{k}
% + g \int_{\epth^-}^{\epth^+} dk \, ( v_k \, c_k^\dag d +  v_k^* \, d^\dag c_k) 
 + g \int dk \, ( v_k \, c_k^\dag d +  v_k^* \, d^\dag c_k)  
\label{ham.gen}
\end{equation}
consisting of a discrete state with creation operator $d^\dagger$ coupled to the field operator $c_k^\dagger$ through the coupling potential $g v_k$.  The discrete state has energy $\epsilon_d$ while the continuum is given by the dispersion $\epsilon_k$, which we suppose extends from a lower threshold 
$\epth^-$ to an upper threshold $\epth^+$.  Typically, in atomic, molecular, and optical systems we will only have a lower threshold while $\epth^+$ is taken to infinity, while in condensed matter systems both thresholds are finite as they delimit the edges of the conduction band.

For our purposes we tend to focus in this paper on studies of the time evolution when the discrete energy $\epsilon_d$ is in the vicinity of the lower threshold $\epth^-$.  In this case, the upper threshold will generally have negligible impact on the survival probability; hence, for our purposes in this section we might as well simply refer to $\epth \equiv \epth^-$ as `the system threshold' and assume that $\epth^+$ is largely inconsequential (or simply taken to infinity $\epth^+ \rightarrow \infty$).

%%%%%%%%%%%%%%%%%
%%%%sub - SECTION BREAK
%%%%%%%%%%%%%%%%%

\subsection{Generic model: discrete spectrum}\label{sec:gen.spec}

We solve for the discrete spectrum of the full Hamiltonian by calculating the Green's function for the discrete state
\begin{equation}
G_{d,d} (z) = \bra d | \frac{1}{z - H} | d \ket
		  = \frac{1}{z - \epsilon_d - \Sigma (z)} \equiv \frac{1}{\eta(z)} .
\label{gen.G.dd}
\end{equation}
Here $\Sigma (z)$ is the self-energy function that determines how the perturbed system eigenvalues are shifted from the unperturbed energy $\epsilon_d$.  
More precisely we may write $\Sigma (z)$ as
\begin{equation}
\Sigma (z) = \Delta (z) + \Lambda (z)
\label{gen.Sigma.z}
\end{equation}
in which $\Delta (z)$ is analytic for all $z$, meanwhile $\Lambda (z)$ represents the non-analytic component of $\Sigma (z)$; specifically $\Lambda (z)$ is undefined along the branch cut associated with the continuum $\epsilon_k$.  Instead, one must define this function on the two different Riemann sheets associated with the branch cut as $\Lambda^\textrm{I,II} (z) = \Lambda (z \pm i 0^+)$.  Then the bound state solutions $\bar{z}_{\textrm{B},i}$ of the system, for example, are written as solutions of
\begin{equation}
\bar{z}_{\textrm{B},i} - \epsilon_d - \Sigma^\textrm{I} (\bar{z}_{\textrm{B},i}) 
%	= z - \epsilon_d - \Delta (z) - \Lambda^\textrm{I} (z)
	= 0
\label{gen.disp.bound}
\end{equation}
in the first sheet.

Further, we have defined in Eq. (\ref{gen.G.dd}) the dispersion function 
$\eta (z) = z - \epsilon_d - \Sigma (z)$, which yields the discrete spectrum as the solutions of the equation $\eta (z) = 0$.  
For many simple systems we may explicitly obtain $\eta (z)$, which then allows us to write the dispersion relation in polynomial form with the eigenvalues given as the roots of the $n$th order polynomial equation
\begin{equation}
P_n (\bar{z}_i) \equiv S (\bar{z}_i) \eta^\textrm{I}(\bar{z}_i) \eta^\textrm{II}(\bar{z}_i) = 0,
\label{gen.disc.disp}
\end{equation}
with the $n$ resulting eigenvalues $\bar{z}_i (\epsilon_d, g)$ functions of the Hamiltonian parameters $\epsilon_d$ and $g$; for this paper we will usually assume we may vary $\epsilon_d$ while $g$ takes a fixed value and may be repressed in the notation from this point forward.  Note that $S(z)$ in Eq. (\ref{gen.disc.disp}) is simply a $z$-dependent factor that might need to be multiplied through to obtain a polynomial.
Let us further assume that for values $\epsilon_d < \epA$ there exists a single bound state in the spectrum with eigenvalue $\zB (\epsilon_d)$, which satisfies $\zB (\epsilon_d) < \epth$ (since we are dealing with an open system, the other eigenvalues might be any combination of anti-bound states, resonant states or anti-resonant states).  However, if we increase the value of $\epsilon_d$ such that $\epsilon_d \rightarrow \epA$ then consequently the bound state energy approaches the threshold $\zB \rightarrow \epth$.  At the precise value $\epsilon_d = \epA$ the bound state 
reaches the continuum threshold $\zB = \epth$ and is absorbed \cite{Economou}.
Hence the bound state is no longer present in the system for values $\epsilon_d \ge \epA$ (instead it becomes an anti-bound state for $\epsilon_d > \epA$).  Our resulting diagonalized Hamiltonian hence takes the form \cite{OPP01}
\begin{equation}
H = 
	\left\{ \begin{array}{lcccc}
		\zB \tilde{d}_\textrm{B}^\dag \tilde{d}_\textrm{B} 
			+ \int dk \, \epsilon_k \, \tilde{c}_k^\dag \tilde{c}_{k}								&	&	&  	& \ \ \epsilon_d < \epA \\
			&	&	&	& 	\\
		 \int dk \, \epsilon_k \, \tilde{c}_k^\dag \tilde{c}_{k}	&	&  	& 	& \ \ \epsilon_d \ge \epA
	\end{array}
	\right.    .
\label{ham.gen.diag}
\end{equation}
Note that for our purposes in this paper, when we talk about `destabilization of the bound state,' we are referring to the transition from the first line to the second line of this equation.\footnote{Also note that other transitions might occur for values of $\epsilon_d$ much larger or much smaller than $\epA$ (appearance of resonance state, etc.) however we primarily focus on the vicinity of the transition $\epsilon_d \sim \epA$ in this paper.}
%%%%%%FOOTNOTE ABOVE ADDED IN V2.
Finally, for ease in the following calculation, we re-write the bound state eigenvalue in terms of 
$\Delta_Q$, the energy gap between the bound state eigenvalue and the continuum threshold, as
\begin{equation}
\zB (\epsilon_d) = \epth - \Delta_Q .
\label{zB.del.Q}
\end{equation}
%%

%%%%%%%%%%%%%%%%%
%%%%sub - SECTION BREAK
%%%%%%%%%%%%%%%%%

\subsection{Generic model: survival probability}\label{sec:gen.surv.prob}

\begin{figure}
\hspace*{0.05\textwidth}
 \includegraphics[width=0.45\textwidth]{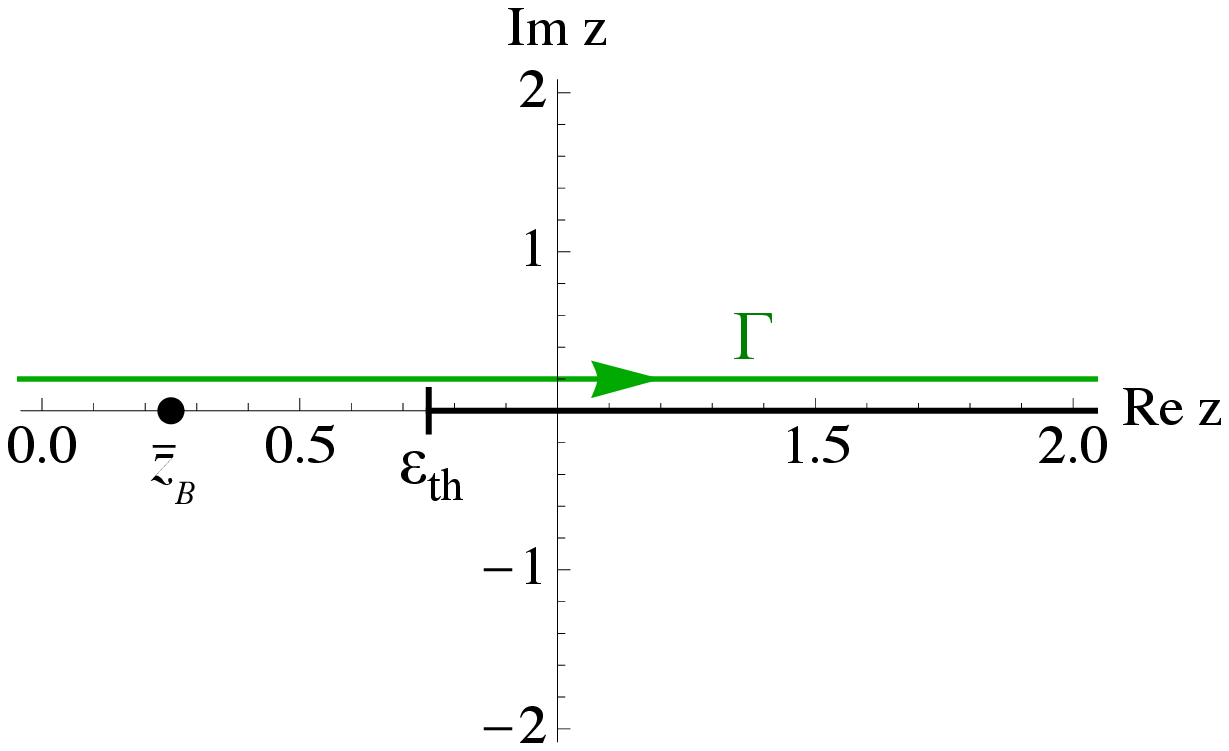}
\hfill
 \includegraphics[width=0.45\textwidth]{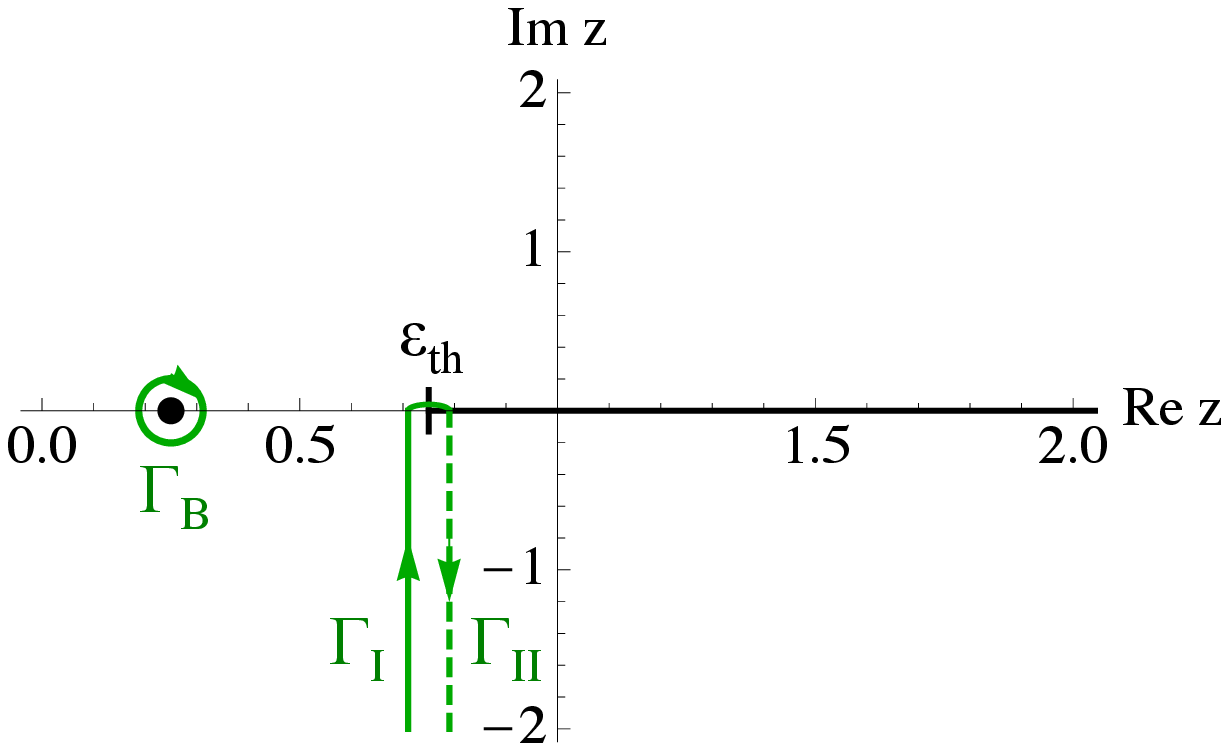}
 \hspace*{0.05\textwidth}
\\
\vspace*{-\baselineskip}
\hspace*{0.08\textwidth}(a)\hspace*{0.465\textwidth}(b)\hspace*{0.4\textwidth}
\\
\vspace*{\baselineskip}
 \caption{
 Complex $z$ plane with the branch cut extending from the threshold $\epth$ to $\infty$ along the real axis.  A single bound state $\bar{z}_\textrm{B}$ appears below the threshold.  The Green line represents the (a) initial contour (Bromwich path), and (b) deformed contour, with unbroken lines in the first sheet and dashed lines in the second sheet.
 }
 \label{fig:gen.contour}
 \end{figure}

Now we calculate the survival probability $P(t) = |A(t)|^2$ in which
\begin{equation}
A(t) = \bra d | e^{-i H t} | d \ket 
	= \frac{1}{2 \pi i} \int_\Gamma dz e^{-i z t} \bra d | \frac{1}{z - H} | d \ket 
	= \frac{1}{2 \pi i} \int_\Gamma dz e^{-i z t} \frac{1}{\eta^I(z)}
\label{gen.surv.amp}
\end{equation}
is the survival amplitude for the system in the prepared initial state $| \Psi \ket = | d \ket$.  Meanwhile, 
$\eta^I (z)$ is the dispersion function defined on the first Riemann sheet and which may be analytically continued into the second sheet on the lower half of the complex energy plane through the branch cut associated with the continuum $\epsilon_k$.
Finally, the contour $\Gamma$ is the Bromwich path \cite{LonghiPRB09,LonghiPRA06} that extends just above the real axis in the first sheet as $[-\infty + i \delta, \infty + i \delta]$ as show in Fig. \ref{fig:gen.contour}(a).

%%%%%%%%%
%%%%%%%%%
%%%%%%%%%
%%%%%%%%%

As a practical means for evaluating the survival amplitude we now deform the contour by the standard approach, which is to drag the entire contour out to infinity in the lower half plane.  In the case 
$\epsilon_d < \epA$ then we will pick up a pole associated with the discrete energy $\zB$ of the bound state located on the real axis below the threshold $\epth$.  The remaining contour consists of a portion at infinity that gives no contribution and a segment $\Gamma_\textrm{I}$ that runs from the threshold 
$\epth$ out to infinity at $\epth - i \infty$.  Meanwhile the segment of the original contour that is initially located just above the continuum must be analytically continued into the second sheet through the branch cut.  As we pull the contour down, we generally pick up a pole for each resonance state that might be present, as well as a second contour $\Gamma_\textrm{II}$ that extends from $\epth$ out to infinity at $\epth - i \infty$ in the second sheet.  Hence ultimately the contributions to the survival amplitude are
\begin{equation}
A(t) = A_\textrm{B} (t) + A_\textrm{th} (t) + A_\textrm{R} (t),
\label{gen.surv.amp.comps}
\end{equation}
in which $A_\textrm{B} (t) = \tilde{R}_\textrm{B} e^{-i \zB t}$ is the contribution from the single bound state with residue $\tilde{R}_\textrm{B}$, $A_\textrm{R} (t)$ is the sum of contributions of any resonance poles (for simplicity in our example, let us assume there are none) and then the contribution from the two contours $\Gamma_\textrm{I}$ and $\Gamma_\textrm{II}$
take the form of the so-called `background integral'
\begin{eqnarray}
A_\textrm{th} (t) 
&	= & \frac{1}{2 \pi i} \int_{\epth - i \infty}^{\epth} dz e^{-i z t} \frac{1}{\eta^\textrm{I}(z)}
		+ \frac{1}{2 \pi i} \int_{\epth}^{\epth - i \infty} dz e^{-i z t} \frac{1}{\eta^\textrm{II}(z)}  \nonumber  \\
&	= &  \frac{1}{2 \pi i} \int_{\epth}^{\epth - i \infty} dz e^{-i z t}
				 \frac{\eta^\textrm{II} (z) - \eta^\textrm{I} (z) }{P_n (z)} S (z)
					  \nonumber  \\
&	= &  \frac{1}{\pi i} \int_{\epth}^{\epth - i \infty} dz e^{-i z t} \frac{\Lambda (z) }{P_n (z)} S (z)
			.
\label{gen.back.int}
\end{eqnarray}
Now, let us explicitly write the polynomial $P_n (z)$ in terms of the single bound state as
\begin{equation}
P_n (z) = (z - \zB) \ R_{n-1} (z)
\label{gen.P.n.zB}
\end{equation}
with $R_{n-1} (z)$ an $n-1$ order polynomial.  Performing a change of the integration variable according to $z = \epth - i y$ we may write the integral Eq. (\ref{gen.back.int}) as
\begin{equation}
A_\textrm{th} (t) 
	= - 2 i e^{-i \epth t} \int_{0}^{\infty} dy \ e^{- y t} 
		\frac{\Lambda (\epth - i y) }{ (\Delta_Q - i y) R_{n-1} (\epth - i y)} .
\label{gen.back.int.y}
\end{equation}
From this point, we make a final coordinate transformation $s = yt$ and re-write the polynomial in the denominator as
\begin{equation}
\bar{R}_{n-1} (\frac{s}{t})
	= R_{n-1} (\epth - i \frac{s}{t}) 
	= \tilde{r}_0 + \tilde{r}_1 \frac{s}{t} + \tilde{r}_2 \frac{s^2}{t^2} + \dots
\label{gen.R.n}
\end{equation}
Finally we obtain the background integral in the standard form as
\begin{equation}
A_\textrm{th} (t) 
	=  \frac{2 e^{-i \epth t}}{it}  \int_{0}^{\infty} ds \ e^{- s} 
		\frac{\Lambda (\epth - i \frac{s}{t}) }{ \Delta_Q \tilde{r}_0 + l_1 \frac{s}{t} + l_2 \frac{s^2}{t^2} + \dots} .
\label{gen.back.int.s}
\end{equation}
in terms of a new polynomial $L_n (s/t) = l_0 + s l_1/t + \dots $ with $l_0 = \Delta_Q \tilde{r}_0$.

Having obtained Eq. (\ref{gen.back.int.s}), we are now in a position to make some comments about the non-Markovian decay.  First, note that for small $t$, the rightmost terms in the denominator of this integral will make the greatest contribution to the integration.  However, we are interested in the long time effects;  let us first focus on the temporal region
\begin{equation}
\frac{l_2}{l_1} \ll t \ll \frac{l_1}{ \Delta_Q \tilde{r}_0} .
\label{gen.cond.near.zone}
\end{equation}
When this condition is satisfied, the second term in the denominator of Eq. (\ref{gen.back.int.s}) will dominate the time evolution.  We may expand the denominator accordingly in order to write the approximation
\begin{equation}
A_\textrm{th} (t) \approx
A_\textrm{th}^\textrm{NZ} (t) 
	=  - \frac{2 e^{-i \epth t}}{i l_1}  \int_{\tau = 1}^{\infty} F(\tau^\prime; t) d \tau^\prime  
\label{gen.back.int.near.zone}
\end{equation}
in which we have defined the function
\begin{equation}
F(\tau; t) 
	= \int_{0}^{\infty} ds \ e^{- s \tau} \Lambda (\epth - i \frac{s}{t})
		,
\label{gen.F.tau.t}
\end{equation}
which will appear again momentarily.  The label ``NZ'' in the above equation stands for long time near zone.

Now let us consider the fully asymptotic case
\begin{equation}
\frac{l_1}{ \Delta_Q \tilde{r}_0} \ll t .
\label{gen.cond.far.zone}
\end{equation}
Here the first term in the denominator of Eq. (\ref{gen.back.int.s}) will dominate the integration, resulting in a new approximation in the form
\begin{equation}
A_\textrm{th} (t) \approx
A_\textrm{th}^\textrm{FZ} (t) 
	=  \frac{2 i e^{-i \epth t}}{t \Delta_Q \tilde{r}_0 }  F(\tau = 1; t).
\label{gen.back.int.far.zone}
\end{equation}
for the long time far zone.

Now let us compare the approximations in the two different zones in Eqs. (\ref{gen.back.int.near.zone}) and (\ref{gen.back.int.far.zone}).  Notice that if we assume
\begin{equation}
\lim_{t \rightarrow \infty} \int_{\tau = 1}^\infty F(\tau^\prime; t) d\tau^\prime  \
	\sim \ \lim_{t \rightarrow \infty} \  F(\tau = 1; t).
\label{gen.F.tau.t.scaling}
\end{equation}
then the two zones obey differing power laws 
$A_\textrm{th}^\textrm{FZ} (t) 
	\sim \frac{1}{t} A_\textrm{th}^\textrm{NZ} (t)$, which in terms of the survival probability itself yields
\begin{equation}
P_\textrm{th}^\textrm{FZ} (t) 
	\sim \frac{1}{t^2} P_\textrm{th}^\textrm{NZ} (t)      \ \ \ \ \ \ \ \ \ \ \ \ \ \ \     \left(  t \gg 1 \right).
\label{gen.FZ.NZ}
\end{equation}
Hence we find a relatively amplified power law decay in the near zone in comparison to the far zone.  However, if we consider the case that $\Delta_Q = 0$ such that the bound state is absorbed into the continuum, then the time scale $T_Q \equiv l_1 / \Delta_Q \tilde{r}_0$ in Eq. (\ref{gen.cond.far.zone}) diverges, and the relatively amplified near zone decay becomes the full asymptotic behavior of the system (that this is true is fairly easy to see from Eq. (\ref{gen.back.int.s}) as well).

%%%%%%%%%%%%%%%%%
%%%%sub - SECTION BREAK
%%%%%%%%%%%%%%%%%

\subsection{Comment on bound state de-stablization in OQS}\label{sec:gen.bound.states}

In this section we comment on the behavior of bound states in open quantum systems and the de-stabilization process itself.  In the previous section we realized a process by which amplification of the long time power law decay may occur in such systems as the bound state is absorbed into the continuum.  Here let us remark on the circumstances under which such a process occurs.

The key point is that this behavior is largely determined by the form of the non-analytic portion of the self-energy $\Lambda(z)$ from Eq. (\ref{gen.Sigma.z}), which contains the branch cut associated with the continuum.  Hence in the single band case when the bound state is absorbed at the edge of the continuum we have $\Lambda(\zB = \epth) = 0$.  Now the form of $\Lambda(z)$ in turn is largely fixed by two pieces from the Hamiltonian: (A) the interaction potential $v_k$, and (B) the continuum density of states $\rho (\epsilon_k)$ (proportional in the 1-D case to the derivative of the inverted dispersion relation $\partial k / \partial \epsilon_k$).  As shown in Ref. \cite{GNHP09}, in the case
\begin{equation}
\lim_{\epsilon \rightarrow \epth} |v_{k(\epsilon)}|^2 \rho (\epsilon) \rightarrow \infty,
\label{van.Hove.state.cond}
\end{equation}
then the condition $\Lambda(\epth) = 0$ can never be satisfied in Eq. (\ref{gen.disp.bound}), which prevents the bound state from ever `touching' the continuum and hence de-stabilization never occurs.  Instead, as $\epsilon_d \gg \epth$ then the bound state will become `stuck' on the outer edge of the continuum, and will only be able to approach the threshold asymptotically (we refer to these type of bound states as {\it persistent bound states} as in Ref. \cite{GNHP09}).

On the other hand, if we consider a system in which 
\begin{equation}
\left[ |v_{k(\epsilon)}|^2 \rho (\epsilon) \right]_{\epsilon= \epth} = 0
\label{bound.state.abs.cond}
\end{equation}
is satisfied, then the condition $\Lambda(\epth) = 0$ may be satisfied for some choice of system parameters at which the bound state will destabilize through absorption into the continuum, ultimately producing an anti-bound state in the second Riemann sheet (or conversely, an anti-bound state may approach the continuum before being ejected as a bound state in the first sheet) \cite{Economou}.  This is equivalent to the localized to virtual state transition studied in Ref. \cite{BCP10}.

Below we will study two models that provide examples of these two cases.  In Sec. \ref{sec:modI} we will consider an infinite chain with a side-coupled impurity, which satisfies the relation in Eq. (\ref{van.Hove.state.cond}).  As a result we find that the persistent bound state may move closer and closer to the band edge, but the condition $\Delta_Q = 0$ cannot be satisfied for finite values of the system parameters.  Hence the time scale in Eq. (\ref{gen.cond.far.zone}) never approaches infinity and the long time far zone behavior can never be entirely eliminated from the system.

However, in Sec. \ref{sec:modII} we study a semi-infinite chain with an endpoint impurity, which satisfies instead the relation Eq. (\ref{bound.state.abs.cond}).  In this case, it is possible to vary the system parameters such that $\Delta_Q = 0$ and then the amplified power law decay associated with the long time near zone will dominate the long time system dynamics even in the asymptotic limit.
As a further point, both bound states and resonance states may be eliminated from the system entirely in this model for a wide range of parameter values (that includes the special point $\Delta_Q = 0$), resulting in non-exponential evolution on all time scales.

%%%%%%%%%%%%%%%%%
%%%%%%%%%% SECTION BREAK
%%%%%%%%%%%%%%%%%

\section{Model realizations of amplified non-Markovian decay}\label{sec:mod}

We now present realizations of the non-Markovian amplification effect relying on two variations of tight-binding chains with coupled impurity sites.  Our two models may be compactly written as
\begin{equation}
H= \epsilon_d d^\dag d
-\frac{t}{2}\sum_{n}(c^\dag_n c_{n+1} + c^\dag_{n+1} c_{n})
+\epsilon_\mu \sum_n c^\dag_n c_n
-g \, (c_{n_0}^\dag d + d^\dag c_{n_0})
\label{ham}
\end{equation}
in which $c_n$ and $d$ are the annihilation operators along the chain and at the
impurity site, respectively, while $t$ sets the tunneling strength along the chain and $g$ gives the coupling strength to the impurity.
Meanwhile $\epsilon_\mu$ gives the chemical potential at each site along the chain and $\epsilon_d$ is the impurity site energy.  We fix the energy scale according to units $t = 1$ and set the origin as 
$\epsilon_\mu = 0$.  Note that the creation operator $c_{n_0}^\dagger$ is the creation operator for the site coupled to the impurity ($n_0$ will be specified below).

\begin{figure}
\hspace*{0.05\textwidth}
 \includegraphics[width=0.4\textwidth]{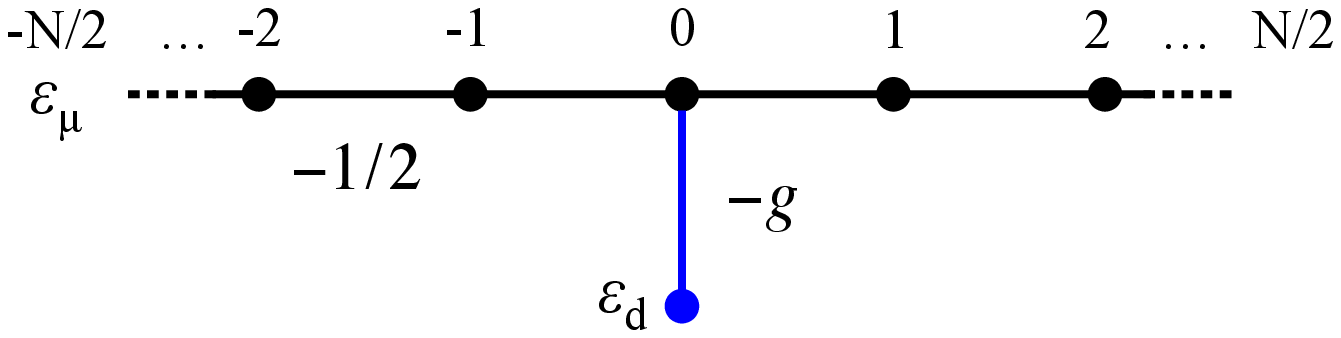}
\hfill
 \includegraphics[width=0.4\textwidth]{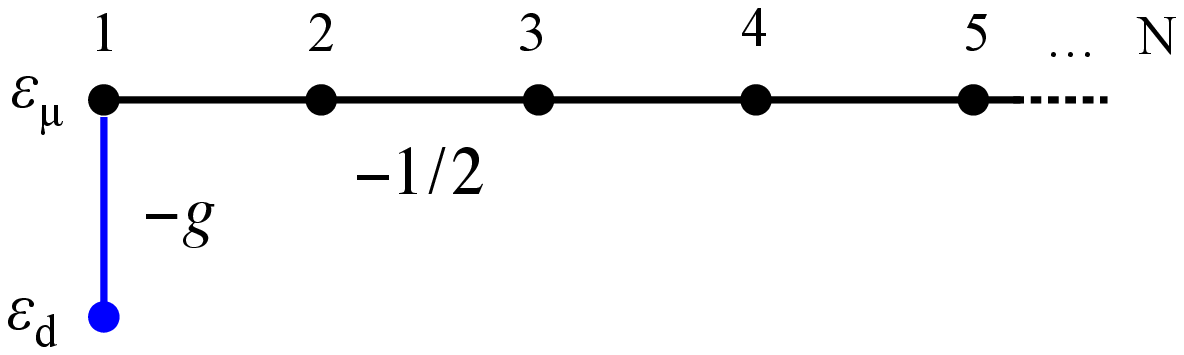}
 \hspace*{0.05\textwidth}
\\
\vspace*{-\baselineskip}
\hspace*{0.08\textwidth}(a)\hspace*{0.465\textwidth}(b)\hspace*{0.4\textwidth}
\\
\vspace*{\baselineskip}
 \caption{(a) Model I: infinite geometry with side-coupled impurity, and (b) Model II: semi-infinite geometry with endpoint impurity.}
% \caption{ }
 \label{fig:model.geo}
 \end{figure}

For {\it Model I}, we have the infinite geometry chain with $N + 1$ sites, indexed as 
$n = -\frac{N}{2}, \dots, \frac{N}{2}$, with periodic boundary conditions and an impurity side-coupled at the middle of the chain $n_0 = 0$ as shown in Fig. \ref{fig:model.geo}(a).
For {\it Model II}, we have the semi-infinite geometry chain with $N$ sites, indexed as 
$n = 1, \dots, \frac{N}{2}$, with open boundary conditions and an end-point impurity attached at the 
$n_0 = 1$ site as shown in Fig. \ref{fig:model.geo}(b).

We visualize that these models may be physically realized as either a semiconductor superlattice as in Ref. \cite{TGP06} or a series of coupled single mode waveguides as in Ref. \cite{LonghiPRL06}.

%%%%%%%%%%%%%%%%%
%%%%sub-SECTION BREAK
%%%%%%%%%%%%%%%%%

\subsection{Model I: persistent bound state supported at threshold by DOS divergence}\label{sec:modI}

Let us now consider Model I, the infinite-chain realization of the Hamiltonian in Eq. (\ref{ham}).  After performing a Fourier transform on the chain sites $c_n$ and taking the limit $N \rightarrow \infty$ we may write
\begin{equation}
H =
\epsilon_d d^\dag d
 + \frac{1}{2}\int_{- \pi}^{\pi} dk \, \epsilon_k \, c_k^\dag c_{k}
 + %\frac{g}{\sqrt{2 \pi}} 
    g \int_{- \pi}^{\pi} v_k dk \, (c_k^\dag d + d^\dag c_k) 
\label{ham.inf}
\end{equation}
in which the dispersion for the diagonalized $c_k$ modes is given by $\epsilon_k = - \cos k$ and the potential $v_k = - 1/\sqrt{2 \pi}$ is simply a constant.  We note immediately the form of the normalized density of states for our dispersion takes the form
\begin{equation}
\rho (\epsilon_k) = \frac{1}{\pi} \frac{1}{\sqrt{1 - \epsilon_k^2}} \propto \frac{\pard k}{\pard \epsilon_k} .
\label{rho.ep}
\end{equation}
with divergences at either band edge $\epsilon_k = \pm 1$.  %As we will see momentarily, these divergences will play a key role in the discrete spectrum for this system by preventing the bound states from being absorbed into the continuum.
Hence at either band edge the condition
\begin{equation}
\lim_{\epsilon \rightarrow \pm 1} |v_{k(\epsilon)}|^2 \rho (\epsilon) \rightarrow \infty
\label{inf.van.Hove.state.cond}
\end{equation}
is satisfied, which is the equivalent of the condition previously appearing in Eq. (\ref{van.Hove.state.cond}).  As a result, the bound state absorption will never be realized in this system, and the amplification of the power law decay will only occur up to a certain timescale.

%%%%%%%%%%%%%%%%%
%%%%sub x 2 SECTION BREAK
%%%%%%%%%%%%%%%%%

\subsubsection{Model I: spectrum}\label{sec:modI.spec}

The discrete spectrum of Model I has been previously studied in Refs. \cite{BCP10,TGP06,HNSP08,Mahan,GNHP09}.
As presented in Ref. \cite{TGP06}, the spectrum may be obtained as the poles of the Green's function for the impurity site, which after re-summation appears as
\begin{equation}
G_{d,d} (z) = \bra d | \frac{1}{z - H} | d \ket
		  = \frac{1}{z - \epsilon_d - \Sigma (z)} \equiv \frac{1}{\eta(z)}
\label{inf.G.dd}
\end{equation}
with the self-energy function $\Sigma (z)$ given by
\begin{equation}
\Sigma (z) \equiv g^2 \int \frac{|v_k|^2}{z - \epsilon_k} dk = \frac{g^2}{\sqrt{z^2 - 1}} .
\label{inf.Sigma.z}
\end{equation}
Note there is only one term in the self-energy, such that $\Lambda (\epsilon) = g^2 / \sqrt{z^2 - 1}$
and $\Delta (z) = 0$.  Further we see immediately that $\Lambda (\pm 1) = 0$ can never be satisfied, as a direct result of Eq. (\ref{inf.van.Hove.state.cond}).

\begin{figure}
\hspace*{0.05\textwidth}
 \includegraphics[width=0.4\textwidth]{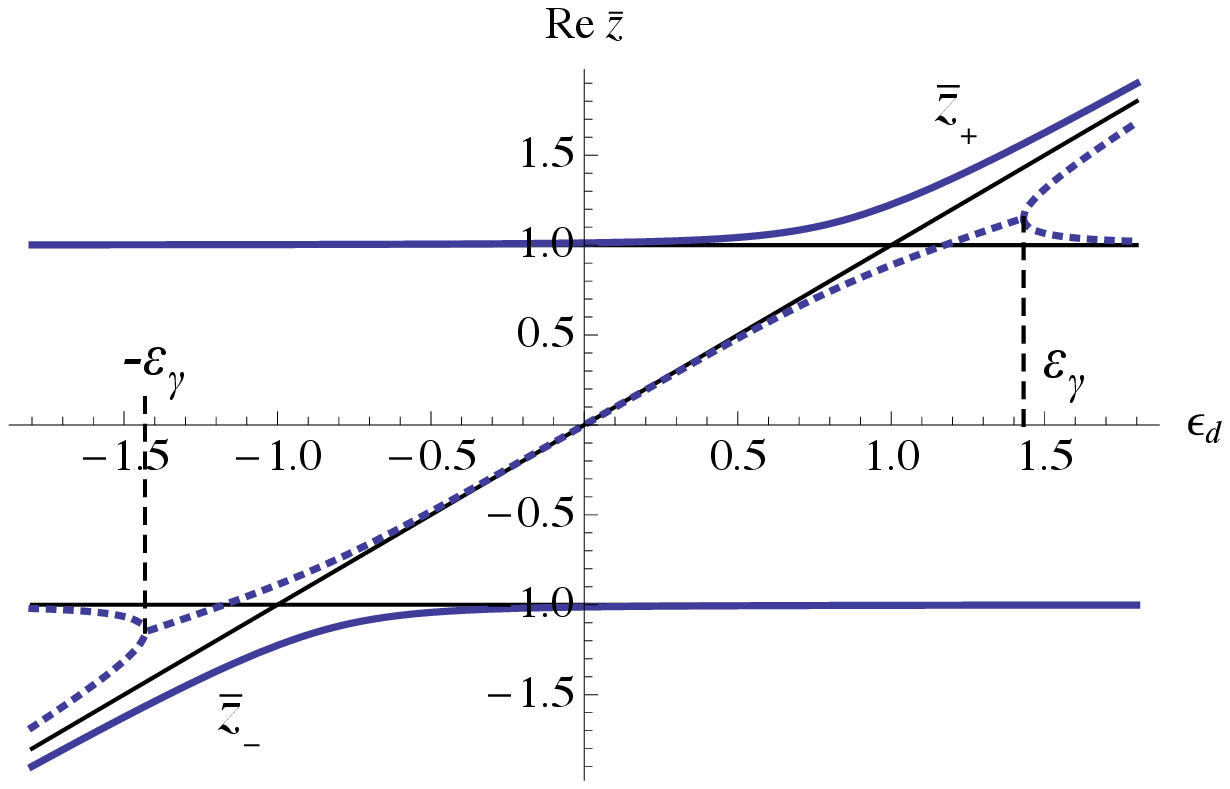}
\hfill
 \includegraphics[width=0.4\textwidth]{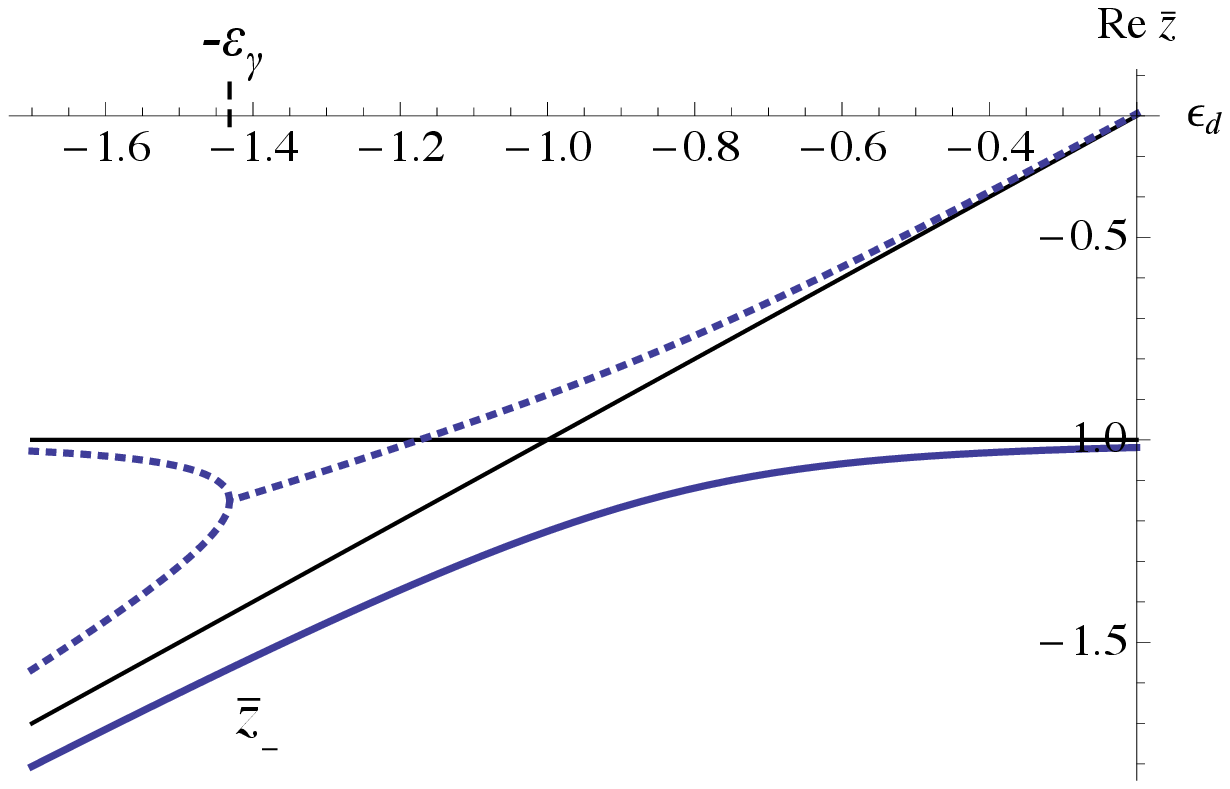}
 \hspace*{0.05\textwidth}
\\
\vspace*{-\baselineskip}
\hspace*{0.08\textwidth}(a)\hspace*{0.465\textwidth}(b)\hspace*{0.4\textwidth}
\\
\vspace*{\baselineskip}
 \caption{(a) Discrete spectrum for Model I as a function of $\epsilon_d$ with $g=0.4$, and (b) close-up in the vicinity of the lower band edge.  The two bound state solutions $\bar{z}_\pm$ are both shown as full, unbroken (blue) curves while anti-bound states and resonance states in the second Riemann sheet are shown with dotted curves.  The lower and upper band edges at $y = \pm 1$ as well as the unperturbed impurity energy $y = \epsilon_d$ are shown with unbroken (black) lines.  The values for 
 $\pm \epsilon_\gamma$ where the resonance state pair forms are also indicated (see Eq. (\ref{inf.e.gamma}) for approximate value).
 }
 \label{fig:modI.spec}
 \end{figure}

The discrete spectrum is now fixed by the poles of the Green's function as the solutions to $\eta(z) = 0$ or
\begin{equation}
z - \epsilon_d = \frac{g^2}{\sqrt{z^2 - 1}} .
\label{inf.disc.disp}
\end{equation}
By squaring Eq. (\ref{inf.disc.disp}) we immediately obtain an equivalent quadratic polynomial equation for the spectrum.  We plot the resulting real part of the spectrum as a function of $\epsilon_d$ in Fig. 
\ref{fig:modI.spec}(a), in which solid lines represent the two bound state solutions appearing in the first Riemann sheet and dashed lines give the two other solutions appearing in the second sheet.

For our purposes, let us focus on the region of the spectrum in the lower half of Fig. \ref{fig:modI.spec}(a), in which the impurity energy lies fairly close to the lower band edge $\epsilon_d \sim -1$.  We show this region in greater detail in Fig. \ref{fig:modI.spec}(b).  For the region in which $\epsilon_d < -\epsilon_\gamma$ (the approximate value of $\epsilon_\gamma$ is given momentarily), the spectrum consists of two bound states, one appearing in the lower half of the spectrum with discrete energy 
$\bar{z}_- \sim \epsilon_d$, and a second on the positive half of the spectrum lying close to the upper band edge $\bar{z}_+ \sim 1$, as well as two anti-bound states in the second Riemann sheet.  When we calculate the survival probability here, not surprisingly we find that the lower bound state $\bar{z}_-$ dominates the time evolution while the upper bound state has a largely negligible effect.

If we choose to gradually increase the value of the parameter $\epsilon_d$, we find that eventually two new features occur in the spectrum.  First, at a specific value of the impurity energy $\epsilon_d = 
- \epsilon_\gamma$, indicated in Fig \ref{fig:modI.spec}, the two anti-bound states collide on the second sheet 
%%%%%%%%%%%%
%%%FOOTNOTE BREAK
to form a resonance, anti-resonance pair\footnote{The resonance, anti-resonance pair forms at a real-valued {\it exceptional point} \cite{Kato,HH01,Moiseyev80},  a process which is described in greater detail in Ref. \cite{GRHS}; for a clever experimental demonstration of this process applying simple LCR circuits, see Ref. \cite{LCR_circuit11}. } with complex conjugate eigenvalues.
%%%FOOTNOTE BREAK
%%%%%%%%%%%%%
This complex conjugate pair persists for all values $|\epsilon_d| < \epsilon_\gamma$ with the approximate value
\begin{eqnarray}
  \epsilon_\gamma = 1 + \frac{3}{2} g^{4/3} + \mathcal O(g^{8/3})
\label{inf.e.gamma}
\end{eqnarray}
obtained in Ref. \cite{TGP06} (a plot of the decay width may also be found in that paper).  
Second, as we continue to increase $\epsilon_d$ 
such that it exceeds values on the order of $\epsilon_d \gtrsim -1 - g^{4/3}$,
the lower bound state takes on a value $\bar{z}_- \lesssim -1$ as it gets `stuck' on the lower band edge, forming a persistent bound state.  
%%%%%%%%%%%%
%%%FOOTNOTE BREAK
\footnote{The transition to the persistent bound state appearing at the band edge is described in detail in Ref. \cite{GNHP09}.}
%%%FOOTNOTE BREAK
%%%%%%%%%%%%%
That this happens can be seen intuitively from Eq. (\ref{inf.disc.disp}); if $z^2 - 1 \sim g^4$ is satisfied in the root on the RHS, we see that both sides of the equation may become order unity.  More precisely, by following the perturbative calculation detailed in Ref. \cite{GNHP09} we obtain
\begin{eqnarray}
\bar{z}_- & = & -1 - \frac{1}{2 (1 + \epsilon_d)^2 } g^4 + \mathcal O(g^8)
						\nonumber \\
		& \equiv & -1 - \tilde{\Delta}_Q + \mathcal O(g^8)
,
\label{inf.eps.van.hove}
\end{eqnarray}
for all values $\epsilon_d \gg -1$.  Notice here that $\bar{z}_-$ never actually reaches to the band edge at $-1$ except in the limit $\epsilon_d \rightarrow \infty$ (including higher order terms would be consistent with this statement).  Hence, for the present model, the bound state is always maintained, hovering just outside the edge of the continuum, as we predicted from the condition reported in 
Eq. (\ref{inf.van.Hove.state.cond}).  We have also defined $\tilde{\Delta}_Q$ here as the gap at lowest order between the bound state and the lower band edge.
%%%Added in v2:
As we evaluate the survival probability below, we find that the background integral naturally appears written in terms of $\tilde{\Delta}_Q$, rendering unnecessary an explicit factorization of the polynomial dispersion in terms of the bound state eigenvalue (as performed previously in Sec. \ref{sec:gen.surv.prob}).

%%%%%%%%%%%%%%%%%
%%%%sub x 2 SECTION BREAK
%%%%%%%%%%%%%%%%%

\subsubsection{Model I: Survival probability}\label{sec:modI.surv.prob}

We now calculate the survival probability along the lines of our previous calculation in Sec. \ref{sec:gen.surv.prob}.  Following the steps leading from Eq. (\ref{gen.surv.amp}) to Eq. (\ref{gen.back.int}) we obtain an expression for the background integral contribution to the survival amplitude for Model I as
\begin{equation}
A_\textrm{th} (t)
	=  \frac{g^2}{\pi i} \int_{-1}^{-1 - i \infty} dz \ e^{-i z t} \ 
		 \frac{\sqrt{z^2 - 1}}{(z - \epsilon_d)^2 (z^2 - 1) - g^4}
	.
\label{inf.back.int}
\end{equation}
Next we make the transformation $z = -1 - i y$ in which $y$ runs from $0$ to $\infty$, followed by the second transformation $s = y t$ in order to obtain
\begin{equation}
A_\textrm{th} (t)
	=  \frac{i e^{i t}}{\pi g^2 t^2} \int_{0}^{\infty} ds \ e^{- s} \ 
		 \frac{\sqrt{s^2 - 2 i t s}}
		 	{1 - i s \frac{T_Q}{t} + s^2 \frac{T_Q T_2}{t^2}
			+ i s^3 \frac{T_Q T_3^2}{t^3} - s^4 \frac{T_Q T_4^3}{t^4}}
	.
\label{inf.back.int.s}
\end{equation}
Here we have defined the key time-scale
\begin{equation}
T_Q \equiv
	\frac{2 (1 + \epsilon_d)^2}{g^4}
		= \frac{1}{\tilde{\Delta}_Q}
\label{inf.T.Q}
\end{equation}
as well as $T_2 \equiv (5 + \epsilon_d)/2 (1 + \epsilon_d)$, 
$T_3 = \sqrt{2 + \epsilon_d}/(1 + \epsilon_d)$, and $T_4 \equiv \left( 2 (1 + \epsilon_d) \right)^{-2/3}$.  Now in the case of $T_2 \ll t \ll T_Q$, we have
\begin{eqnarray}
A_\textrm{th}^\textrm{NZ} (t)
&	\approx & - \frac{e^{i t}}{\pi g^2 t T_Q} \int_{0}^{\infty} ds \ e^{- s} \ 
		 \frac{\sqrt{s^2 - 2 i t s}}{s}					\nonumber  \\
	%%%%
&	= & - \frac{e^{i t}}{\pi g^2 t T_Q} \int_{\tau = 1}^{\infty} \bar{F} \left( \tau^\prime, t  \right) d \tau^\prime
											\nonumber  \\
	%%%%
&	= & \frac{ g^2 e^{it}}{2 \sqrt{\pi} (1 + \epsilon_d)^2 t} \Psi (-1/2, 0 ; - 2 i t)
\label{inf.back.int.near.zone}
\end{eqnarray}
in which $\Psi (a, c ; z)$ is the confluent hypergeometric function and
\begin{equation}
\bar{F}(\tau ; t) 
	\equiv \int_0^\infty ds \ e^{- s \tau} \sqrt{s^2 - 2 i t s}
	= \frac{i t e^{-i t \tau} K_1 (-it \tau)}{\tau}
	,
\label{inf.T.Q}
\end{equation}
with $K_1 (x)$ the modified Bessel function.
As we generally expect the condition $T_2 \ll t \ll T_Q$ will apply for $t \gg 1$ we apply the asymptotic expansion for the confluent hypergeometric function ($\Psi (a, c; z) \rightarrow z^{-a}$ for $z \rightarrow \infty)$ to obtain
\begin{equation}
A_\textrm{th}^\textrm{NZ} (t)
	\approx \frac{e^{3 \pi i / 4} g^2 e^{it} }{\sqrt{2 \pi} (1 + \epsilon_d)^2 t^{1/2}}.
\label{inf.back.int.near.zone.asymp}
\end{equation}
Hence we find the survival probability goes like $P^\textrm{NZ} (t) \sim t^{-1}$ in the long time near zone.  
%%%%%%
Meanwhile from Eq. (\ref{inf.back.int.s}) in the long time far zone $t \gg T_Q$ we utilize the asymptotic behavior $K_1 (x) \rightarrow \sqrt{\pi / 2 x} e^{-x}$ to obtain
\begin{eqnarray}
A_\textrm{th}^\textrm{FZ} (t)
&	\approx & - \frac{i e^{i t}}{\pi g^2 t^2} \bar{F}(\tau ; t) 			\nonumber  \\
	%%%%
&	= & - \frac{1}{\pi g^2 t} K_1 (-i t)
		,
\label{inf.back.int.near.zone}
\end{eqnarray}
which in the asymptotic limit gives
\begin{equation}
A_\textrm{th}^\textrm{FZ} (t)
	\approx \frac{e^{\pi i / 4} e^{it} }{\sqrt{2 \pi} g^2 t^{3/2}}
	.
\label{inf.back.int.far.zone.asymp}
\end{equation}
%%
%%
%%%%FIGURE%%%%
%%%%%%%%%%%%
\begin{figure}
\hspace*{0.25\textwidth}
 \includegraphics[width=0.5\textwidth]{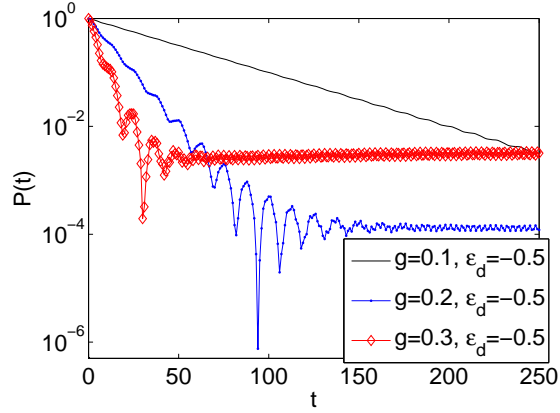}
\hfill
 \hspace*{0.05\textwidth}
\\
%\vspace*{-\baselineskip}
%\hspace*{0.08\textwidth}(a)\hspace*{0.465\textwidth}(b)\hspace*{0.4\textwidth}
%\\
%\vspace*{\baselineskip}
 \caption{Numerical results for the time evolution $P(t)$ for Model I for the three cases.
}
 \label{fig:modI.surv.prob}
 \end{figure}
 %%%%%%%%%%%%%
 %%%%END FIGURE%%%

Using Eqs. (\ref{inf.back.int.near.zone.asymp}) and (\ref{inf.back.int.far.zone.asymp}) to calculate
the survival probabilities $P_\textrm{th}^\textrm{NZ} = |A_\textrm{th}^\textrm{NZ} (t)| \sim t^{-1}$ and $P_\textrm{th}^\textrm{FZ} = |A_\textrm{th}^\textrm{FZ} (t)| \sim t^{-3}$, we immediately verify that the predicted relation Eq. (\ref{gen.FZ.NZ}) holds true for the present model.
Unfortunately however, the fact that both bound states and a resonant state are present in the system for the parameter choices we are interested in (i.e. $\Delta_Q$ small) results in a decay profile in which it is difficult to clearly distinguish the time evolution behavior in the near zone; this is demonstrated by numerical results in Fig. \ref{fig:modI.surv.prob}.
%%%Added in v2:
For the case $g = 0.1, \epsilon_d = -0.5$ (large red diamonds) we see that the exponential decay dominates the visible time evolution up to about $t < 20$, then the bound state is the dominant feature around $t > 50$.  In the approximate region $20 < t < 50$ the evolution evolves according to interference between these two terms, resulting in an oscillation pattern.  Then for $g = 0.2,  \epsilon_d = -0.5$ (blue dots) the decay width (proportional to $g^2$) is smaller resulting in an elongated timescale $t < 60$ for the exponential decay.  Finally for $g = 0.1,  \epsilon_d = -0.5$ the decay width is even smaller, resulting in an exponential decay that outlives the plot parameters.

We see that the power law decay effects are not apparent in any of these plots.  Fortunately, we may overcome this difficulty for certain parameter choices in the case of Model II.

%%%%%%%%%%%%%%%%%
%%%%%sub- SECTION BREAK
%%%%%%%%%%%%%%%%%

\subsection{Model II: bound state absorbed into continuum}\label{sec:modII}

We now turn to the Model II realization of the Hamiltonian Eq. (\ref{ham}), consisting of a semi-infinite-chain with an endpoint impurity.  We perform a half-range Fourier transform on the chain sites $c_n$ and take the limit $N \rightarrow \infty$ in order to write
\begin{equation}
H =
\epsilon_d d^\dag d
 + \frac{1}{2}\int_{0}^{\pi} dk \, \epsilon_k \, c_k^\dag c_{k}
 + %\frac{g}{\sqrt{2 \pi}} 
    g \int_{0}^{\pi} v_k dk \, (c_k^\dag d + d^\dag c_k) 
\label{ham.inf}
\end{equation}
with the same tight-binding dispersion $\epsilon_k = - \cos k$ as for Model I, hence the density of states takes precisely the same form as Eq. (\ref{rho.ep}), including the square root form of the divergences at either band edge $\epsilon_k = \pm 1$

Meanwhile the potential takes the form $v_k = - \sin k /\sqrt{\pi}$, which introduces some quite different properties into the system in comparison to Model I.  The primary point is that
at either band edge the condition
\begin{equation}
\left[ |v_{k(\epsilon)}|^2 \rho (\epsilon) \right]_{\epsilon = \pm 1} = 0
\label{zero.van.Hove.state.cond}
\end{equation}
is satisfied, which is the equivalent of the condition Eq. (\ref{bound.state.abs.cond}) that allows for a bound state to become destabilized as it touches either edge of the conduction band at specific parameter values.  As we study the spectrum in greater detail in the next section we will also present the precise value $\epsilon_d = \epA$ where this destabilization occurs (assuming a fixed value of $g$).

%%%%%%%%%%%%%%%%%
%%%%sub x 2 SECTION BREAK
%%%%%%%%%%%%%%%%%

\subsubsection{Model II: spectrum}\label{sec:modII.spec}

The discrete spectrum and the power law decay for Model II has been studied previously in 
Ref. \cite{LonghiPRL06} for the case of $\epsilon_d = 0$ (all sites with equal energy) while only the coupling to the endpoint may be varied.
For our purposes of illustrating a simple generic feature we will focus instead on the particular case of a fixed endpoint coupling ($g = 1/2$ in Fig. \ref{fig:model.geo} (b)) while the impurity energy 
$\epsilon_d$ may be varied.
The power law decay has been studied in models similar to these in Refs. 
\cite{thesis,DBP08,LonghiPRA06}.  
Just as for Model I, we obtain the discrete spectrum for the present model from the Green's function at the endpoint impurity site as in Eq. (\ref{inf.G.dd}).  And as in Eq. (\ref{inf.Sigma.z}), it is  a straight-forward exercise to obtain the self-energy function for arbitrary $g$ as
\begin{equation}
\Sigma (z) = 2 g^2 \left( z - \sqrt{z^2 - 1} \right)
\label{semi.Sigma.z}
\end{equation}
for the semi-infinite chain model, in which it is now possible for $\Lambda(\pm1) = 0$ to be satisfied for 
$\Lambda (z) = 2 g^2 \sqrt{z^2 - 1}$.

%%%%%%%%%%
%%%%%%%%%%
\begin{figure}
\hspace*{0.05\textwidth}
 \includegraphics[width=0.44\textwidth]{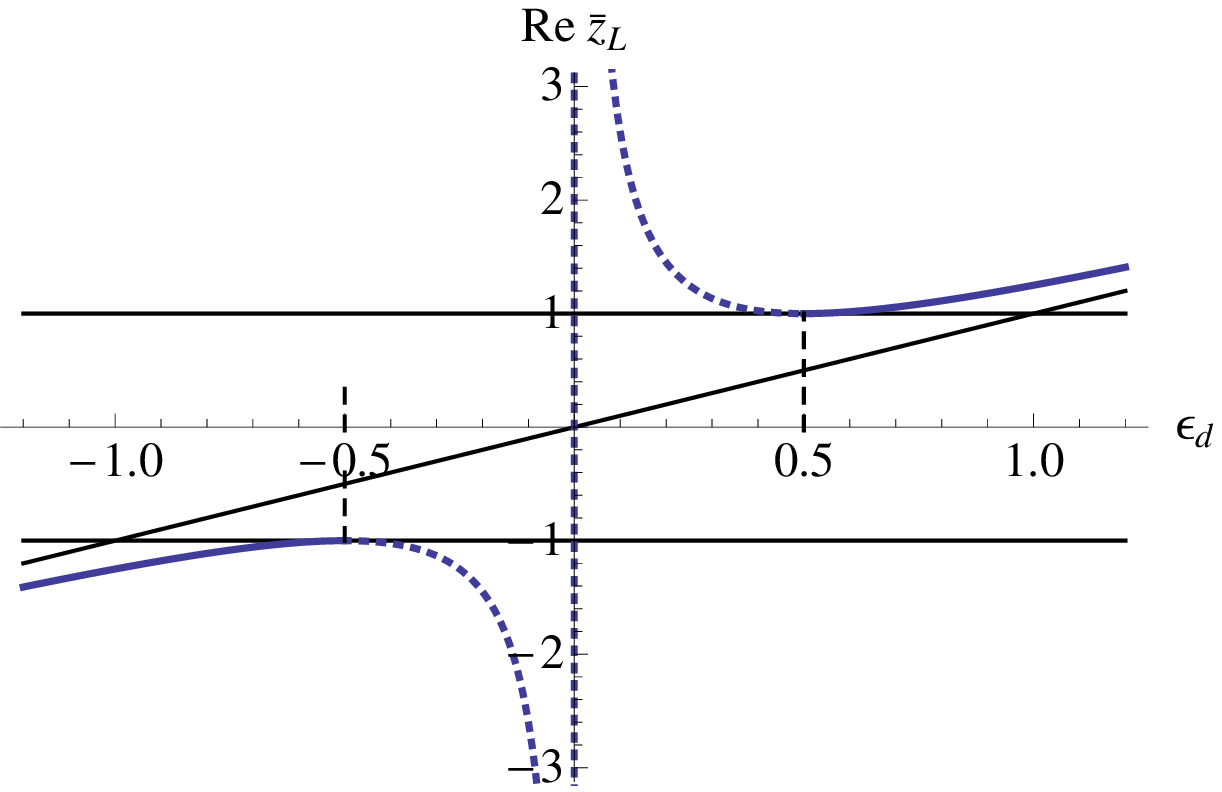}
\hfill
 \includegraphics[width=0.4\textwidth]{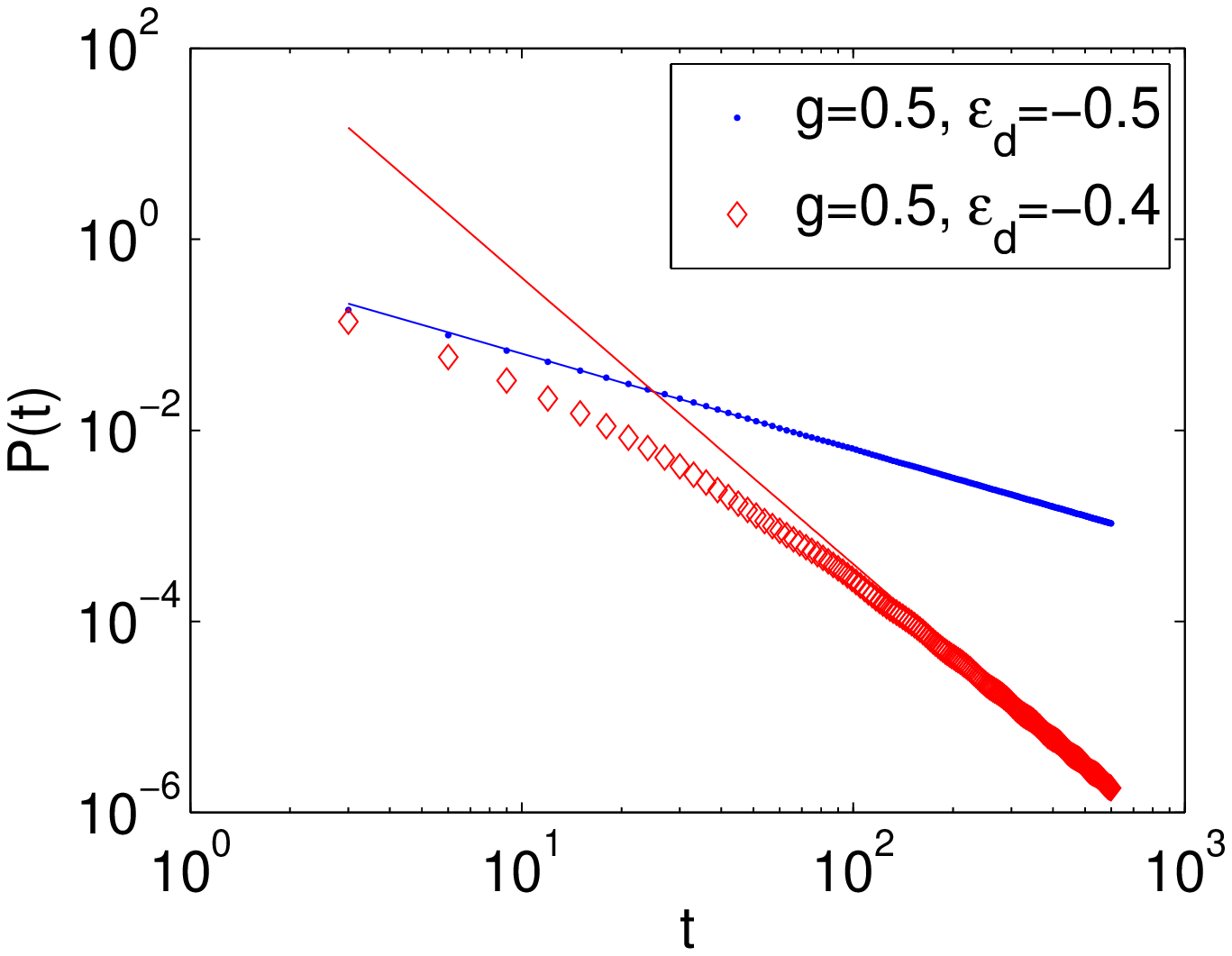}
 \hspace*{0.05\textwidth}
\\
\vspace*{-\baselineskip}
\hspace*{0.08\textwidth}(a)\hspace*{0.465\textwidth}(b)\hspace*{0.4\textwidth}
\\
\vspace*{\baselineskip}
 \caption{(a) Discrete spectrum for Model II as a function of $\epsilon_d$ with $g=1/2$ (linear case).  The lower and upper band edges as well the unperturbed energy $y = \epsilon_d$ are all shown with solid black lines.  The bound state in the first sheet is shown by a full, unbroken (blue) curve while the anti-bound state in the second sheet is represented with a (blue) dotted curve.  Vertical dashed lines represent the point $\epsilon_d = \pm 1/2$ at which the bound state brushes the continuum to form an anti-bound state (or vice versa).  (b)  Numerical results for the time evolution $P(t)$ for Model II for two cases: $\epsilon_d = -0.4$ (large red diamonds) along with the red line representing 
 the far zone evolution $P^{FZ} (t) \sim t^{-3}$, and $\epsilon_d = -0.5$ (small blue dots) along with the blue line representing the near zone evolution $P^{NZ} (t) \sim t^{-1}$.}
 \label{fig:modII.spec}
 \end{figure}
%%%%%%%%%%
%%%%%%%%%%

The discrete spectrum is obtained from $\eta (z) = z - \epsilon_d - \Sigma (z) = 0$ as
\begin{equation}
z - \epsilon_d = \frac{1}{2} \left( z - \sqrt{z^2 - 1} \right)
\label{semi.disc.disp}
\end{equation}
for the specific case $g = 1/2$.
In the more general case of arbitrary $g$, we would obtain an equivalent quadratic polynomial after squaring out the root on the RHS of this equation.  
However, for our present case we instead find a simple linear equation.  Before presenting the solution, we first note there is a special point in parameter space associated with the values $z = \pm 1$ at which the root in Eq. (\ref{semi.disc.disp}) vanishes.  What happens at this special point is that a bound state appearing in the first Riemann sheet touches the edge of the continuum, before giving rise to an anti-bound state in the second sheet (or vice-versa) \cite{DBP08,Economou}.  Plugging the values 
$z = \pm 1$ into Eq. (\ref{semi.disc.disp}) yields $\epsilon_d = \pm 1/2$ as the precise points in parameter space where these transitions occur.  It can be further demonstrated that for $| \epsilon_d | > 1/2$ there is exactly one bound state present in the system, and for $| \epsilon_d | < 1/2$ there is exactly one anti-bound state.  In either case, the linear dispersion yields the single solution as
\begin{equation}
\bar{z}_\textrm{L} (\epsilon_\textrm{d})
	=  \epsilon_d + \frac{1}{4 \epsilon_d},
\label{semi.z.L}
\end{equation}
%%
%We immediately obtain the solutions to the quadratic as
%%
%\begin{equation}
%\bar{z}_\pm (\epsilon_\textrm{d}, g^2)
%	=  \frac{ \epsilon_\textrm{d} (1 - g^2) 
%			\pm g^2 \sqrt{\epsilon_\textrm{d}^2 - \left(1-2g^2 \right)} }{1 - 2g^2}.
%\label{semi.z.pm}
%\end{equation}
%%
which holds for the bound and anti-bound state alike.  We plot the spectrum as a function of $\epsilon_d$ in the Fig. \ref{fig:modII.spec}(a), in which unbroken curves show the bound state portion of the spectrum, while the dotted curves show the anti-bound state.  It's interesting to note that in the anti-bound case $| \epsilon_d | < 1/2$, there is neither a bound state nor a resonance state in the system, hence the time evolution is non-exponential on all time scales \cite{LonghiPRL06,thesis}.

%%%%%%%%%%%%%%%%%
%%%%sub x 2 SECTION BREAK
%%%%%%%%%%%%%%%%%

\subsubsection{Model II: Survival probability}\label{sec:modII.surv.prob}

Again we write the survival probability for Model II by following the standard method from Sec. \ref{sec:gen.surv.prob} to obtain
\begin{eqnarray}
A_\textrm{th} (t)
&	= &  - \frac{1}{2 \pi i \epsilon_d} \int_{-1}^{-1 - i \infty} dz \ e^{-i z t} \ 
		 \frac{\sqrt{z^2 - 1}}{z - \bar{z}_\textrm{L}}        \nonumber \\
&	= &  \frac{e^{i t}}{2 \pi i \epsilon_d \Delta_Q t^2} \int_{0}^{\infty} ds \ e^{- s} \ 
		 \frac{\sqrt{s^2 - 2 i t s}}
		 	{ 1 + i s \frac{T_Q}{t} }
		,
\label{semi.back.int}
\end{eqnarray}
following the usual transformations $z = -1 - i y$ then $s = y t$.  Here $\Delta_Q \equiv \zL + 1$ is the gap between the bound state and the lower band edge at $z = -1$ and $T_Q = (\Delta_Q)^{-1}$ is the usual time scale dividing the near zone and far zone.  In the long time near zone $1 \ll t \ll T_Q$ we have
\begin{eqnarray}
A_\textrm{th}^\textrm{NZ} (t)
&	\approx & - \frac{i e^{i t}}{2 \pi i \epsilon_d t} \int_{\tau = 1}^{\infty} \bar{F} 
				\left( \tau^\prime, t  \right) d \tau^\prime
											\nonumber  \\
	%%%%
&	= & - \frac{ e^{it}}{2 \sqrt{\pi} \epsilon_d t} \Psi (-1/2, 0 ; - 2 i t)
											\nonumber   \\
	%%%%
&	\approx & - \frac{e^{3 \pi i / 4} e^{it} }{ \sqrt{2 \pi} \epsilon_d t^{1/2}}
\label{semi.back.int.near.zone}
\end{eqnarray}
with the usual $t^{-1/2}$ behavior.
Meanwhile as we turn to the long time far zone in the fully asymptotic case $t \gg T_Q$, we find from Eq. (\ref{semi.back.int}) that
\begin{eqnarray}
A_\textrm{th}^\textrm{FZ} (t)
&	\approx & - \frac{1}{2 \pi \epsilon_d \Delta_Q t}  K_1(- i t)
											\nonumber  \\
	%%%%
&	\approx & - \frac{e^{\pi i / 4} e^{it} }{ 2 \sqrt{2 \pi} \epsilon_d \Delta_Q t^{3/2}}
	.
\label{semi.back.int.far.zone}
\end{eqnarray}
Clearly we see that in the case that the bound state touches the continuum at $\epsilon_d = 1/2$ ($\Delta_Q = 0$), then $T_Q \rightarrow \infty$ and the long time near zone gives the dominant behavior even in the asymptotic limit.

We illustrate this by numerical results for $P(t)$ in Fig. \ref{fig:modII.spec}(b); 
in the case that $\epsilon_d = -0.4$ (red diamonds), there is no bound state present in the system and we clearly observe the transition from the near zone with slope coefficient 
$\sim -1$ (blue line) and the far zone with coefficient $\sim -3$ (red line).  Meanwhile for the value 
$\epsilon_d = -0.5$ (small blue dots) we have realized the case $\Delta_Q = 0$, hence the 
far zone time scale has been pushed out and the near zone behavior dominates the evolution even on larger time scales.

%%%%%%%%%%%%%%%%%
%%%%%%%%%% SECTION BREAK
%%%%%%%%%%%%%%%%%

\section{Conclusion}\label{sec:conc}

In this paper we have demonstrated in simple terms that there is some connection between the proximity of a bound state to threshold and the precise details of the non-exponential decay of the survival probability at long times in open quantum systems.  This connection is encapsulated in Eq. (\ref{gen.back.int.s}) in which it is clear that closing the gap $\Delta_Q$ between the threshold and bound state should have some effect on the long time power law decay (most likely, it will be enhanced).  
Further the existence of the gap gives rise to a time scale that divides the long time system dynamics into two zones: the long time near zone and the long time far zone.  The question of how the power law decay will compare in these two zones largely depends on whether the condition in Eq. (\ref{gen.F.tau.t.scaling}) holds.

In Sec. \ref{sec:mod} we explored this question for two specific nearest neighbor models with attached impurity levels.  In these two cases Eq. (\ref{gen.F.tau.t.scaling}) indeed held, and hence the power law decay was enhanced in the long time near zone in comparison to the far zone according to 
$P^\textrm{FZ} \sim \frac{1}{t^2} P^\textrm{NZ}$ as predicted in Eq. (\ref{gen.FZ.NZ}).  Further, for Model I the divergent van Hove singularity in the density of states at the edge of the continuum prevented the bound state from being able to reach the band edge and closing the gap $\Delta_Q$ (that is to say, there was a persistent bound state as defined in Ref. \cite{GNHP09}).
Hence one could only approach the regime in which the near zone dominates the power law decay for very large times only in the limit $\epsilon_d \rightarrow \infty$, in which case the decay contribution in Eq. (\ref{inf.back.int.near.zone.asymp}) would vanish anyways.  However, for Model II the precise form of the interaction potential essentially cancels out the divergence from the density of states, which results in a spectrum in which it is possible to close the gap $\Delta_Q = 0$ for precise values of the system parameters.  Doing so gives rise to a situation where the enhanced power law decay associated with the long time near zone pushes out the far zone behavior entirely and dominates the time evolution in the fully asymptotic limit.

Interestingly, it has also been demonstrated in open quantum systems that the same crossover from a bound state to an anti-bound state at the continuum threshold gives rise to a maximum in the local density of states response function at the impurity site \cite{BCP10}.  Indeed, this result is closely related to the amplification of the power law decay at threshold \cite{DBP08}.

However, the precise details of the transitions in the asymptotic decay behaviors for a generic system remains an open question.  On the one hand, the $t^{-3}$ behavior in the far zone appears to be a rather generic feature \cite{Sudarshan}.
On the other hand, one would hope to obtain a more general understanding of the near zone behavior.
One might gain insight into this question by considering what more general statements might be made about the condition expressed 
in Eq. (\ref{gen.F.tau.t.scaling}) and under what conditions one might expect this relation to hold.
One interesting proposal would be to study the non-exponential decay dynamics in a two-dimensional system, in which the inverse square root singularities ubiquitous in one-dimensional systems would be replaced with logarithmic divergences.  This will be a future direction of study.

Finally for completeness we note that even the typical exponential decay itself experiences anomalous behavior in the vicinity of the threshold.  As has been pointed out in quite different physical contexts, Fermi's golden rule, which predicts that the decay rate to lowest order approximation is proportional to the density of states, breaks down in the vicinity of a singularity in the density of states \cite{TGP06,GNHP09,PTG05}.

\begin{acknowledgement}
We thank Ingrid Rotter, Satoshi Tanaka and E. C. G. Sudarshan for insightful discussions.  The research of S. G. was supported by CQIQC, the Sloan research fellowship of D. S. and MPI-PKS.
\end{acknowledgement}

%The style of the following references should be used in all documents.

\end{document}